\documentclass{ws-procs9x6}

\usepackage{epsfig}

\newcommand{\mpi}{m_\pi}
\newcommand{\gev}{\, {\rm GeV}}
\newcommand{\mev}{\, {\rm MeV}}
\newcommand{\fm}{\, {\rm fm}}

\begin{document}

\title{Quenched Chiral Physics in Baryon Masses}

\author{R.~D. YOUNG, D.~B. LEINWEBER and A.~W. THOMAS}

\address{	Special Research Centre for the
		Subatomic Structure of Matter,
		and Department of Physics and Mathematical Physics,
		University of Adelaide, Adelaide SA 5005,
		Australia\\
E-mail: ryoung@physics.adelaide.edu.au; dleinweb@physics.adelaide.edu.au; athomas@physics.adelaide.edu.au}

\author{S.~V. WRIGHT}

\address{	Division of Theoretical Physics,
		Department of Mathematical Sciences,\\
		University of Liverpool,
		Liverpool L69 3BX, U.K.\\
E-mail: svwright@amtp.liv.ac.uk}

%%%%%%%%%%%%%%%%%%%%%%%%%%%%%%%%%%%%%%%%%%%%%%%%%%%%%%%%%%%%%%
 
\maketitle

\abstracts{
Recent work has identified that the primary differences between
quenched and dynamical spectroscopy can be described by chiral loop
effects. Here we highlight the features of this study.
}

\vspace{-10cm} \hfill \mbox{\small ADP-02-99/T537, LTH564}
\vspace{9cm}

\section{Introduction}
Chiral symmetry has long been known to play an important role in the
low energy properties of QCD. Theoretical studies of the
non-perturbative features of QCD, expressed in terms of the fundamental
theory, have proven to be most successful in the field of lattice
gauge theory. Lattice studies are typically restricted to relatively
large values of the $u$ and $d$ masses, where chiral effects are
highly suppressed. As a consequence, direct observation of chiral
properties is a challenge in lattice simulations. The study of
low-lying baryon spectroscopy in both quenched and full QCD has
provided a direct connection between lattice results and chiral
physics\cite{Young:2001nc,Young:2002cj}.

The fact that one is restricted to quark masses much larger than the
physical values means that, in addition to all the usual
extrapolations (e.g. to the infinite volume and continuum limits), if
one wants to compare with empirical hadron observables, one must also
have a reliable method of extrapolation to the chiral limit. Any such
extrapolation must incorporate the appropriate chiral corrections,
arising from Goldstone boson loops, which give rise to rapid,
non-linear variations as the chiral limit is approached. The
importance of incorporating such behaviour has been successfully
demonstrated for a number of hadronic observables, see
Ref.~[3] for a review and the references
contained therein.

The quenched approximation is a widely used tool for studying
non-perturbative QCD within numerical simulations of lattice gauge
theory. With an appropriate choice of the lattice scale and at
moderate to heavy quark masses, this approximation has been shown to
give only small, systematic deviations from the results of full QCD
with dynamical fermions. Although no formal connection has been
established between full and quenched QCD, the similarity of the
results has led to the belief that the effects of quenching are small
and hence that quenched QCD provides a reasonable approximation to the
full theory\cite{Aoki:1999yr}. Under a more reliable choice of lattice
scale, where chiral effects are negligible, clear differences are
observed between quenched and dynamical results\cite{Bernard:2001av}.

%%%%%%%%%%%%%%%%%%%%%%%%%%%%%%%%%%%%%%%%%%%%%%%%%%%%%%%%%%%%%%%%%%%%%%%%%%%

\section{Chiral Physics and the Quenched Approximation}

Chiral symmetry is spontaneously broken in the ground state of QCD. As
a consequence, the pion is a Goldstone boson characteristic of this
broken symmetry. The pion mass then behaves as $\mpi^2 \propto m_q$,
the well known Gell-Mann--Oakes--Renner (GOR) relation. In principle,
this relation is only guaranteed for quark masses near zero. Explicit
lattice calculations show that it holds over an enormous range, as
high as $m_\pi \sim 1\gev$. These almost massless Goldstone bosons
couple strongly to low-lying baryon states --- particularly the
nucleon ($N$) and delta ($\Delta$).

Meson-loop diagrams involving Goldstone bosons coupling to baryons
give rise to non-analytic behaviour of baryon properties as a function
of quark mass. At light quark masses these corrections are large and
rapidly varying. At heavy quark masses these contributions are
suppressed and hadron properties are smooth, slowly varying functions
of quark mass. The scale of this transition is characterised by the
inverse size of the pion-cloud source. Below pion masses of about
$400$--$500\mev$ these non-analytic contributions become increasingly
important, while they rapidly become negligible above this point.
Since lattice simulations are typically restricted to the domain where
$\mpi\gtrsim 500\mev$, these rapid-varying chiral effects must be
incorporated phenomenologically.

\begin{figure}[!t]
\begin{center}
{\epsfig{file=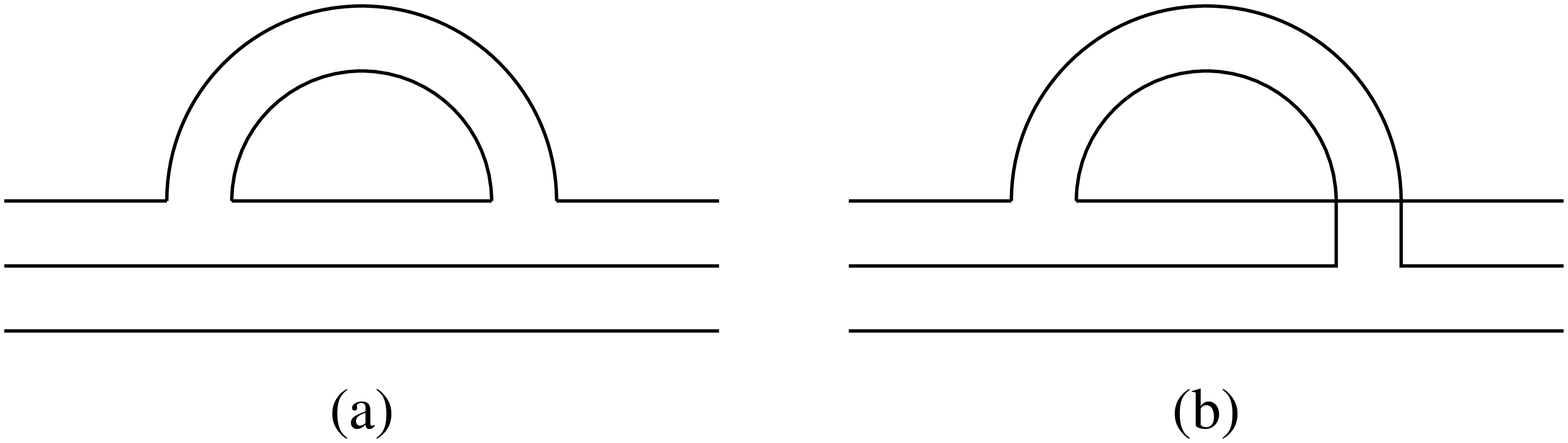, width=6cm, angle=0}}
\caption{Quark flow diagrams of pion loop contributions
appearing in QCD.
\label{fig:piQF}}
\end{center}
\end{figure}

Within the quenched approximation dynamical sea quarks are absent from
the simulation. As a consequence the structure of meson-loop
contributions is modified. In the physical theory of QCD, meson-loop
diagrams can be described by two topologically differing types. A
typical meson-loop diagram may be decomposed into those where the loop
meson contains a sea quark, such as Fig.~\ref{fig:piQF}(a), and those
where the loop meson is comprised of pure valence quarks, see
Fig.~\ref{fig:piQF}(b). These diagrams involving the sea quark, type
(a), are obviously absent in the quenched approximation and
consequently only a subset of the contributions to the physical theory
are included.  This type of argument, together with SU(6) symmetry,
is precisely that described for the evaluation of non-analytic
contributions to baryon magnetic moments by
Leinweber\cite{Leinweber:2001jc}.  This quark flow approach is
analogous to the original approach to chiral perturbation theory for
mesons performed by Sharpe\cite{Sharpe:1990me,Sharpe:1992ft}.

In addition to the usual pion loop contributions, quenched QCD (QQCD)
contains loop diagrams involving the flavour singlet $\eta'$ which
also give rise to important non-analytic structure. Within full QCD
such loops do not play a role in the chiral expansion because the
$\eta'$ remains massive in the chiral limit.  On the other hand, in
the quenched approximation the $\eta'$ is also a Goldstone
boson\cite{Sharpe:1992ft} and the $\eta'$ propagator has the same
kinematic structure as that of the pion.

\begin{figure}[!t]
\begin{center}
{\epsfig{file=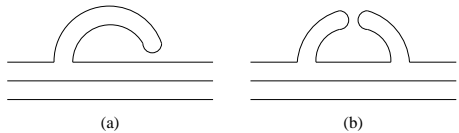, width=6cm, angle=0}}
\caption{Quark flow diagrams of chiral $\eta'$ loop contributions
appearing in QQCD:  
(a) axial hairpin, (b) double hairpin.
\label{fig:etaQF}}
\end{center}
\end{figure}

As a consequence there are two new chiral loop contributions unique to
the quenched theory. The first of these corresponds to an axial
hairpin diagram such as that indicated in Fig.~\ref{fig:etaQF}(a).
This diagram gives a contribution to the chiral expansion of baryon
masses which is non-analytic at order $\mpi^3$. The second of these new
$\eta'$ loop diagrams arises from the double hairpin vertex as
pictured in Fig.~\ref{fig:etaQF}(b). This contribution is particularly
interesting because it involves two Goldstone boson propagators and is
therefore the source of more singular non-analytic behaviour -- linear
in $\mpi$. In studying the extrapolation of quenched lattice results
it is essential to treat these contributions on an equal footing to
the pion-loop diagrams discussed earlier.
%

%%%%%%%%%%%%%%%%%%%%%%%%%%%%%%%%%%%%%%%%%%%%%%%%%%%%%%%%%%%%%%%%%%%%%%%%%%%

\section{Chiral Extrapolations}
In general, the coefficients of the leading (LNA) and next-to leading
non-analytic (NLNA) terms in a chiral expansion of baryon masses are
very large. For instance, the LNA term for the nucleon mass is $\delta
m_N^{(\rm LNA)} = -5.6\, m_\pi^3$ (with $m_\pi$ and $\delta m_N^{(\rm
LNA)}$ in GeV). With $m_\pi = 0.5\gev$, quite a low mass for current
simulations, this yields $\delta m_N^{(\rm LNA)} =0.7\gev$ --- a huge
contribution. Furthermore, in this region hadron masses in both full
and quenched lattice QCD are found to be essentially linear in
$m_\pi^2$ or equivalently quark mass, whereas $\delta m_N^{(\rm LNA)}$
is highly non-linear. The challenge is therefore to ensure the
appropriate LNA and NLNA behaviour, {\it with the correct
coefficients}, as $m_\pi \rightarrow 0$, while making a sufficiently
rapid transition to the linear behaviour of actual lattice data, where
$\mpi$ becomes large.

A reliable method for achieving all this was proposed by
Leinweber~{\it et~al.}\cite{Leinweber:1999ig}
They fit the full (unquenched) lattice data with the form:
\begin{equation}
M_B = \alpha_B + \beta_B \mpi^2 + \Sigma_B (\mpi, \Lambda) ,
\label{eq:FullFit}
\end{equation}
where $\Sigma_B$ is the total contribution from those pion loops which
give rise to the LNA and NLNA terms in the self-energy of the
baryon. The extension to the case of quenched QCD is achieved by
replacing the self-energies, $\Sigma_B$, of the physical theory by the
corresponding contributions of the quenched theory\cite{Young:2002cj},
$\tilde\Sigma_B$.

The linear term of Eq.~(\ref{eq:FullFit}), which dominates for $m_\pi
\gg \Lambda$, models the quark mass dependence of the pion-cloud
source --- the baryon without its pion dressing. This term also serves
to account for loop diagrams involving heavier mesons, which have much
slower variation with quark mass.

The diagrams for the various meson-loop contributions are evaluated
using a phenomenological regulator. This regulator has the effect of
suppressing the contributions as soon as the pion mass becomes
large. At light quark masses the self-energies, $\Sigma_B$, provide
the same non-analytic behaviour as $\chi$PT, independent of the choice
of regulator. Therefore the functional form, Eq.~(\ref{eq:FullFit}),
naturally encapsulates both the light quark limit of $\chi$PT and the
heavy quark behaviour observed on the lattice.

\begin{figure}[!t]
\begin{center}
\mbox{
\vspace*{5mm}$ \Sigma_N = $
\hspace{1mm}
\centering{\
        \epsfig{angle=0,figure=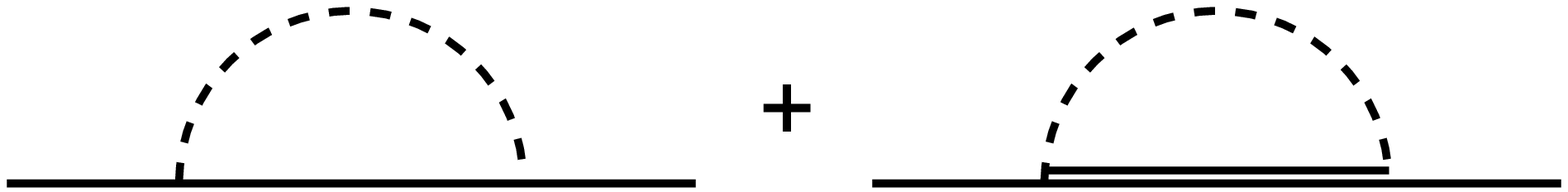, width=7cm} }
}
\mbox{\put(-30,20){\large$\pi$}\put(-140,20){\large$\pi$}}

\vspace*{5mm}

\mbox{
\vspace*{5mm}$ \Sigma_\Delta = $
\hspace{1mm}
\centering{\
        \epsfig{angle=0,figure=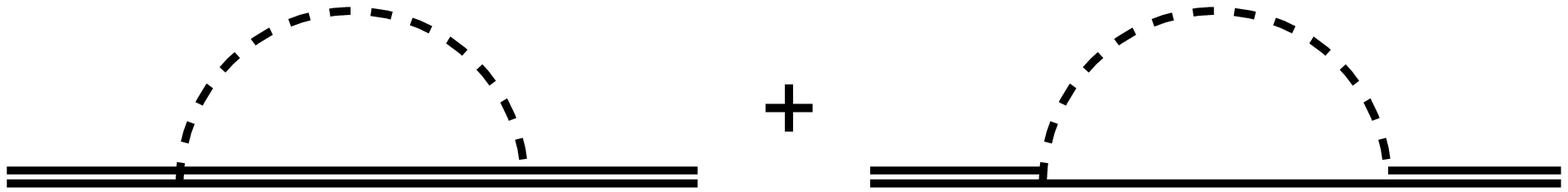, width=7cm} }
}
\mbox{\put(-30,20){\large$\pi$}\put(-140,20){\large$\pi$}}
\caption{Illustrative view of the meson-loop self-energies,
$\Sigma_B$, in full QCD. These diagrams give rise to the LNA and NLNA
contributions in the chiral expansion. Single (double) lines denote
propagation of a $N$ ($\Delta$).
\label{fig:fullSE}}
\end{center}
\end{figure}

\begin{figure}[!t]
\begin{center}
\mbox{
\put(-30,35){$\tilde\Sigma_N = $}
\hspace{1mm}
\centering{\
        \epsfig{angle=0,figure=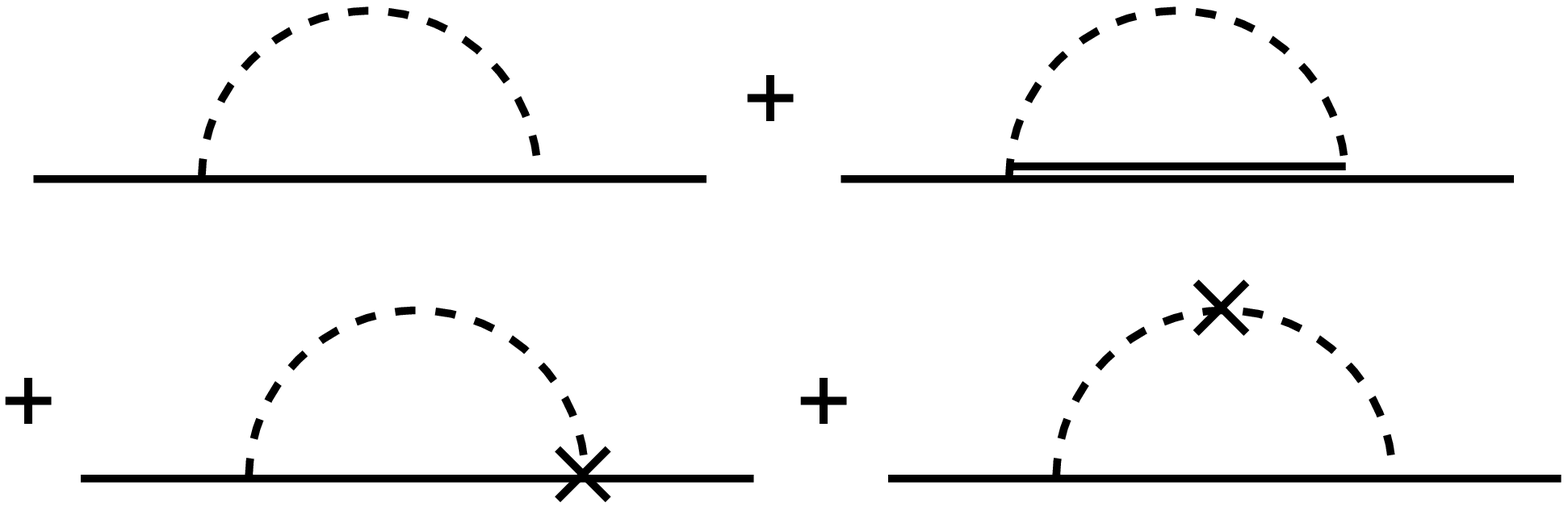, width=7cm} }
}
\mbox{\put(-140,60){\large$\pi$}}\put(-40,60){\large$\pi$}
\mbox{\put(-135,20){\large$\eta'$}}\put(-30,20){\large$\eta'$}
\caption{Illustrative view of the meson-loop self-energies,
$\Sigma_B$, in quenched QCD. These diagrams give rise to the LNA and
NLNA contributions in the chiral expansion. A cross represents a
hairpin vertex in the $\eta'$ propagator. Single (double) lines denote
propagation of a $N$ ($\Delta$).
\label{fig:quenchSE}}
\end{center}
\end{figure}

We consider the leading order diagrams containing only the lightest
Goldstone degrees of freedom. These are responsible for the most rapid
non-linear variation as the quark mass is pushed down toward the
chiral limit. In the physical theory we consider only those diagrams
containing pions, as depicted in Fig.~\ref{fig:fullSE}.  In QQCD there
exist modified pion loop contributions and the additional structures
arising from the $\eta'$ behaving as a Goldstone boson. The diagrams
contributing to the nucleon self-energy in the quenched approximation
are shown in Fig.~\ref{fig:quenchSE}, the $\Delta$ can be described by
analogous diagrams.

For the evaluation of the quenched quantities we assume that the
parameters of the chiral Lagrangian exhibit negligible differences
between quenched and 3-flavour dynamical simulations. This is a
working hypothesis with no better guidance yet available, but the
successful results of this work demonstrate the self-consistency of
such an assumption\cite{Young:2001nc}.  Only with further accurate
lattice simulations at light masses will one be able to determine the
extent to which our hypothesis holds.

\begin{figure}[!t]
\begin{center}
{\epsfig{file=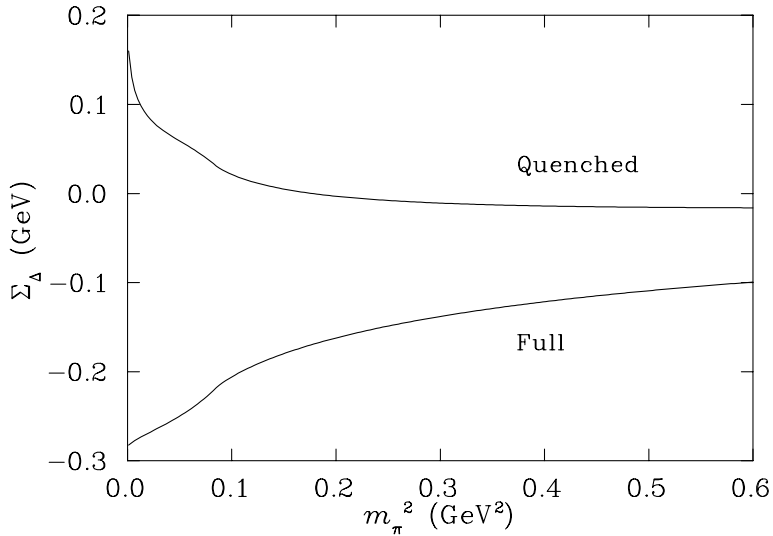, width=7cm, angle=90}}
\caption{Net contributions to the $\Delta$ self-energy evaluated with
dipole regulator at $\Lambda=0.8\gev$.
\label{fig:seD}}
\end{center}
\end{figure}

To highlight the differences in the self-energy contributions we show
the net contributions to the $\Delta$ in full ($\Sigma_\Delta$) and
quenched ($\tilde\Sigma_\Delta$) QCD in Fig.~\ref{fig:seD}. For
details of the breakdown of the individual contributions we refer the
reader to our longer article\cite{Young:2002cj}.  The significant point
to note is that whereas the meson cloud of the $\Delta$ is attractive
in full QCD, it exhibits repulsive behaviour within the quenched
approximation.
%

%%%%%%%%%%%%%%%%%%%%%%%%%%%%%%%%%%%%%%%%%%%%%%%%%%%%%%%%%%%%%%%%%%%%%%%%%%%

\section{Fitting Lattice Data}
The lattice data considered in this analysis comes from the recent
paper of Bernard~{\em et~al.}\cite{Bernard:2001av}  These simulations
were performed using an improved Kogut-Susskind quark action, which
shows evidence of good scaling\cite{Bernard:1999xx}.  Unlike the
standard Wilson fermion action, masses determined at finite lattice
spacing are good estimates of the continuum limit results.

We are particularly concerned with the chiral extrapolation of baryon
masses and how their behaviour is affected by the quenched
approximation. In such a study, it is essential that the method of
scale determination is free from chiral contamination. One such method
involves the static-quark potential. As low-lying pseudoscalar mesons
made of light quarks exhibit negligible coupling to hadrons containing
only heavy valence quarks, the low energy effective field theory plays
no role in the determination of the scale for these systems. In fixing
the scale through such a procedure one constrains all simulations,
quenched, 2-flavour, 3-flavour {\em etc.}, to match phenomenological
static-quark forces.  Effectively, the short range ($0.35\sim 0.5\fm$)
interactions are matched across all simulations.

A commonly adopted method involving the static-quark potential is the
Sommer scale\cite{Sommer:1993ce,Edwards:1997xf}.  This procedure
defines the force, $F(r)$, between heavy quarks at a particular length
scale, namely $r_0\simeq0.5\, {\rm fm}$. Choosing a narrow window to
study the potential avoids complications arising in dynamical
simulations where screening and ultimately string breaking is
encountered at large separations. The lattice data analysed in this
report uses a variant of this definition, choosing to define the force\cite{Bernard:2001av}
at $r_1=0.35\, {\rm fm}$ via $r_1^2 F(r_1)=1.00$.

The non-analytic chiral behaviour is governed by the infrared regions
of the self-energy integrals. Due to the finite volume of lattice
simulations much of this structure will not be captured. For this
reason we evaluate the self-energy corrections with pion momenta
restricted to those available on the particular
lattice\cite{Young:2002cj,Leinweber:2001ac}. In this way we get a
first estimate of the discretisation errors in the meson-loop
corrections. In no way does this account for any artefacts associated
with the pion-cloud source.

Current lattice data is insufficient to reliably extract the dipole
regulator parameter, $\Lambda$. We fit all data choosing a common
value to describe all vertices, $\Lambda=0.8\gev$. This choice has
been optimised\cite{Young:2002cj} to highlight the main result of this
analysis. We note that the value of $\Lambda$ which we find is
indeed consistent with phenomenological estimates which
suggest that this should be somewhat less than $1\gev$.

\begin{figure}[!t]
\begin{center}
\epsfig{file=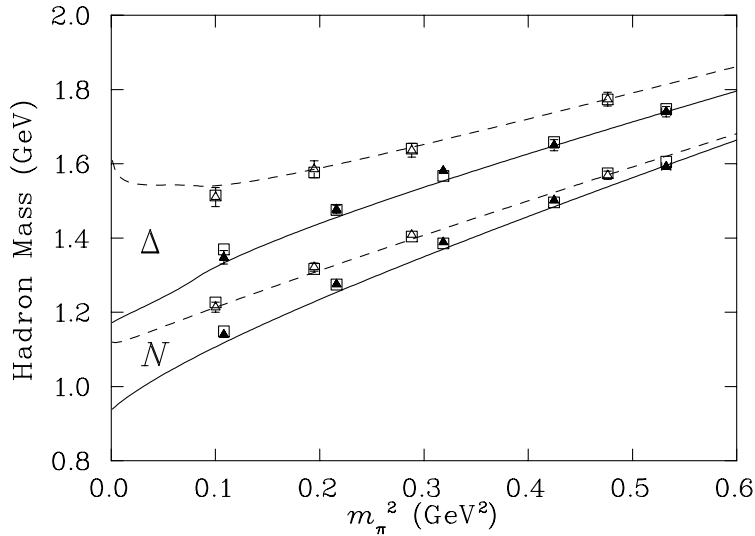, width=7cm, angle=90}
\caption{Fit (open squares) to lattice
data\protect\cite{Bernard:2001av} (Quenched $\vartriangle$, Dynamical
$\blacktriangle$) with adjusted self-energy expressions accounting for
finite volume and lattice spacing artifacts.  The infinite-volume,
continuum limit of quenched (dashed lines) and dynamical (solid lines)
are shown. The lower curves and data points are for the nucleon and
the upper ones for the $\Delta$.}
\label{fig:fqFit}
\end{center}
\end{figure}

We fit both quenched and dynamical simulation results to the form of
Eq.~(\ref{eq:FullFit}) with appropriate discretised self-energies.
These fits are shown in Fig.~\ref{fig:fqFit}. It is the open squares
which should be compared with the lattice data. These points
correspond to evaluation of the self-energies on the discretised
momentum grid. The lines represent a restoration of the continuum
limit in the self-energy evaluation. Discrepancies between the
continuum and discrete version only become apparent at light quark
masses, this corresponds to the Compton wavelength of the pion
becoming comparable to the finite spatial extent of the lattice.

The important result of this study is that the behaviour of the
pion-cloud source is found to be quite similar in both quenched and
dynamical simulations. Once the self-energies corresponding to the
given theory are incorporated into the fit, the linear terms are found
to be in excellent agreement. Our best fit parameters, for the
selected dipole mass $\Lambda=0.8\gev$, are shown in
Table~\ref{tab:fitparams}.
\begin{table}[!t]
\tbl{Best fit parameters for both full and quenched data sets with
dipole regulator, $\Lambda=0.8\gev$. All masses are in 
GeV.\vspace*{1pt}}
{\begin{tabular}{|lcccc|}
\hline
Simulation	& $\alpha_N$	& $\beta_N$	& $\alpha_\Delta$	& $\beta_\Delta$     \\
\hline
Physical	& $1.27(2)$     & $0.90(5)$	& $1.45(3)$		& $0.74(8)$          \\
Quenched	& $1.24(2)$     & $0.85(6)$	& $1.45(4)$		& $0.72(11)$         \\
\hline 
\end{tabular}
\label{tab:fitparams}}
\end{table}
Here we observe the remarkable agreement between quenched and
dynamical data sets for $N$ and $\Delta$ masses over a wide range of
pion mass. This leads to the interpretation that the primary effects
of quenching can be described by the modified chiral structures which
give rise to the LNA and NLNA behaviour of the respective theories.

The success of fitting the $N$ and $\Delta$ data sets with a common
regulator lends confidence to an interpretation of the mass splitting
between these states. Examination of the self-energy contributions in
full QCD suggests that only about $50\mev$ of the observed $300\mev$
$N$-$\Delta$ splitting arises from pion loops\cite{Young:2002cj}.  The
dominant contribution to the hyperfine splitting would then naturally
be described by some short-range quark-gluon interactions.

%%%%%%%%%%%%%%%%%%%%%%%%%%%%%%%%%%%%%%%%%%%%%%%%%%%%%%%%%%%%%%%%%%%%%%%%%%%

\section{Conclusions}
We have demonstrated the strength of fitting lattice data with a
functional form which naturally interpolates between the domains of
heavy and light quarks. The extrapolation formula gives a reliable
method for the extraction of baryon masses at realistic quark
masses. Although the quenched approximation gives rise to more
singular behaviour in the chiral limit, these contributions are
quickly suppressed with increasing pion mass. Within the quenched
approximation only limited curvature is observed for the $N$ down to
low quark masses. In contrast, we find some upward curvature of the
$\Delta$ mass in the light quark domain.

The observation that the source of the meson cloud has remarkably
similar behaviour within both quenched and physical simulations is of
considerable importance. One can describe the primary effects of
quenching by the meson-loop contributions which give rise to the most
rapid, non-linear variation at light quark masses. This leads one to
the possibility of applying this result to obtain more physical
results from quenched simulations. The structure of the meson-cloud
source can be determined from quenched simulations and then the chiral
structures of the physical theory can be incorporated
phenomenologically. Natural extension of this work leads to the
analysis of further hyperons to investigate the applicability over a
range of particles.

%%%%%%%%%%%%%%%%%%%%%%%%%%%%%%%%%%%%%%%%%%%%%%%%%%%%%%%%%%%%%%%%%%%%%%%%%%%

\end{document}